\documentclass[twocolumn,pre,aps,showpacs,preprintnumbers,amsmath,amssymb,superscriptaddress]{revtex4-1}
\usepackage{epstopdf}
\usepackage{graphicx}
\usepackage{graphicx,epsf} 
\usepackage{dcolumn}
\usepackage{bm}
\usepackage{latexsym}
\usepackage{color}

\begin{document}

\title{Competing Effects of Social Balance and Influence}

\author{P. Singh}\thanks{Present address: Northwestern Institute on Complex Systems, Northwestern University, 600 Foster Street, Evanston, IL 60208-4057 USA}
\affiliation{Department of Physics, Applied Physics, and Astronomy, Rensselaer Polytechnic Institute,
110 8$^{th}$ Street, Troy, NY, 12180-3590 USA}
\affiliation{Social Cognitive Networks Academic Research Center, Rensselaer Polytechnic Institute, 110 8$^{th}$ Street, Troy, NY, 12180-3590 USA}

\author{S. Sreenivasan}
\affiliation{Department of Physics, Applied Physics, and Astronomy, Rensselaer Polytechnic Institute, 110 8$^{th}$ Street, Troy, NY, 12180-3590 USA}
\affiliation{Social Cognitive Networks Academic Research Center, Rensselaer Polytechnic Institute, 110 8$^{th}$ Street, Troy, NY, 12180-3590 USA}
\affiliation{Department of Computer Science, Rensselaer Polytechnic Institute, 110 8$^{th}$ Street, Troy, NY, 12180-3590 USA}

\author{B. K. Szymanski}
\affiliation{Social Cognitive Networks Academic Research Center, Rensselaer Polytechnic Institute, 110 8$^{th}$ Street, Troy, NY, 12180-3590 USA}
\affiliation{Department of Computer Science, Rensselaer Polytechnic Institute, 110 8$^{th}$ Street, Troy, NY, 12180-3590 USA}
\affiliation{Faculty of Computer Science and Management, Wroclaw University of Technology, 50-370 Wroclaw, Poland}

\author{G. Korniss}
\affiliation{Department of Physics, Applied Physics, and Astronomy, Rensselaer Polytechnic Institute, 110 8$^{th}$ Street, Troy, NY, 12180-3590 USA}
\affiliation{Social Cognitive Networks Academic Research Center, Rensselaer Polytechnic Institute, 110 8$^{th}$ Street, Troy, NY, 12180-3590 USA}

\begin{abstract}
We study a three-state ({\it leftist, rightist, centrist}) model
that couples the dynamics of social balance with an external
deradicalizing field. The mean-field analysis shows that there
exists a critical value of the external field $p_c$ such that for a
weak external field ($p$$<$$p_c$), the system exhibits a metastable
fixed point and a saddle point in addition to a stable fixed point.
However, if the strength of the external field is sufficiently large
($p$$>$$p_c$), there is only one (stable) fixed point which
corresponds to an all-centrist consensus state (absorbing state). In
the weak-field regime, the convergence time to the absorbing state
is evaluated using the quasi-stationary distribution and is found to
be in agreement with the results obtained by numerical simulations.

\end{abstract}

\pacs{
87.23.Ge, 
89.75.Fb, 
02.50.Ey  
}

\maketitle

\section{Introduction}
Structural balance is considered to be one of the key driving
mechanisms of social dynamics \cite{Heider,Wasserman_1994,Lambiotte_PNAS2010} and since
its introduction by Heider~\cite{Heider}, it has been studied
extensively in the context of social networks
\cite{Lambiotte_PNAS2010,Cartwright_1956,Antal_PRE,Antal_Physica,Marvel_balance}. In a
socially interacting population, relationships among individuals
({\it links} in the underlying social network) can be classified as
{\it friendly} (+) or {\it unfriendly} (-). Evolution of these links
is governed by the theory of {\it structural balance}, also referred
to as {\it social balance}. The underlying axioms behind this theory
are: (i) a friend of my friend or an enemy of my enemy is my friend,
and (ii) a friend of my enemy or an enemy of my friend is my enemy.
In the context of social networks, a triangle is said to be
unbalanced if it contains an odd number of unfriendly links
\cite{Antal_PRE,Antal_Physica}. According to the theory of social
balance, these unbalanced triangles have a tendency to evolve to
balanced configurations \cite{Heider,Cartwright_1956}. This might
happen by transitioning nodes and/or interpersonal links in such a
way that the conditions of social balance are satisfied. For
example, a triad in which two mutually antagonistic individuals have
a common friend, is by definition unbalanced but can become balanced
by requiring either the common friend to choose a side or the
mutually antagonistic two nodes to reconcile their conflict and
become friends. A structurally balanced network contains no
unbalanced triangles.
Here, we construct an individual-based model where the dynamics (in
part) is driven by structural balance, but (unlike in previous works
\cite{Antal_PRE,Antal_Physica,Marvel_balance,Galam96,GalamEPJB14,Galam_war_peace}) a change in the state
of an edge is the direct consequence of a change in the state
(opinion) of one of the nodes the edge connects. Another key
feature of the model studied here is that triadic (three-body) interactions
among nodes have been considered as opposed to dyadic (pair-wise) interactions in 
three-state models~\cite{Mobilia2013,Vazquez2004,Zhang_scirep}).

In this paper, we consider a population where each individual is in
one of the three possible opinion states ({\it leftist}, {\it
rightist}, or {\it
centrist})~\cite{Mobilia2013,Vazquez2004,Zhang_scirep}. A link that
connects two {\it extremists} of opposite type (i. e. the link
between a leftist and a rightist) is considered to be unfriendly
while all other links are friendly. Thus, a triangle containing one
node of each type (leftist, rightist, and centrist) is unbalanced.
An unbalanced triangle can be balanced in a number of ways with each
minimal change solution requiring one node in the triangle updating
its opinion. In a model with extremist and moderate opinion states
of individuals, Marvel et al.~\cite{Marvel_PRL2012} showed that
moderation by external stimulus is a way to have a society adopt a
moderate viewpoint (non-social deradicalization). Here, in our
model, we consider a similar external influence field (e.g.,
campaigns, advertisements) which converts extremists into centrists.
Thus, the system is governed by the competing effects of social
balance and influence.

\begin{figure}[!htbp]
\centerline{
\epsfxsize=8 cm
\epsfclipon
\epsfbox{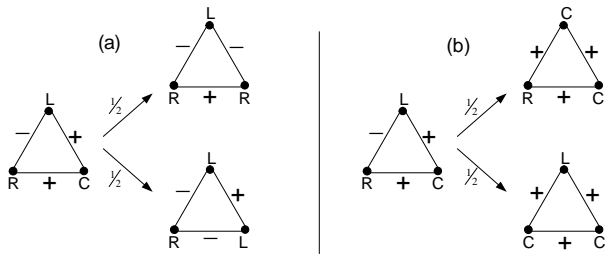}
}
\caption{The balance step is taken with probability $(1-p)$. However, balance can be achieved in two ways:
(a) with probability $\alpha$ a centrist is converted into an extremist and since it can either be a leftist
or a rightist, its flavor is chosen randomly with equal probability $1/2$.
(b) Alternatively, with probability $(1-\alpha)$ an extremist (either leftist or rightist with
equal probability $1/2$) is converted to a centrist.}
\label{Fig1}
\end{figure}

At each time step either with probability $p$ (i) random node is
selected and if it is an extremist, it is converted to a centrist,
or with the complementary probability $(1-p)$ (ii) a random triangle
is selected and if unbalanced, it is balanced by either converting
(with a probability $\alpha$) a centrist into an extremist or [with
a probability $(1-\alpha)$] an extremist into a centrist.
Furthermore, since an extremist can either be a leftist or a
rightist, a choice of converted extremist flavor is made with equal
probability ($\frac{1}{2}$) as shown in Fig.~\ref{Fig1}.
\section{Fully-connected networks (Mean-field analysis)}
\subsection{Fixed points of the system}
For a fully-connected network, at any given time the state of a system of size $N$ can be described
by two numbers - the density (fraction) of leftists ($x$) and the
density of rightists ($y$) - as we can eliminate the density of
centrists ($z$) since $x+y+z=1$. Thus, the evolution can be mapped
onto the $xy$ plane. A finite system will always have only one
absorbing fixed point (for $p > 0$) that is a consensus state where every node has
adopted the centrist opinion. First, for simplicity, we consider
this dynamics on an infinite complete graph where every node is
connected to every other node (i.e., in the mean-field limit). Starting from
an arbitrary state ($x > 0$, $y > 0$, $z > 0$), in the absence
of an external influencing field ($p=0$), the final state of the system
is either polarized ($z=0$) or is a coalition where mixed population of centrists
 and extremists of one kind (either leftist or rightist) coexists.
For $p=0$, the whole triangular boundary of the phase space in the $xy$ plane becomes absorbing, thus, a 
pure consensus state in this case can not be reached through transitions from a different
initial state (see Appendix A).
For $p>0$, the evolution of $x,y$ densities is governed by the following rate
equations:
\begin{eqnarray}
\frac{dx}{dt} &=& -p x  +  3 (2\alpha-1) (1-p) x y (1 - x - y)  \\
\frac{dy}{dt} &=& -p y  +  3 (2\alpha-1) (1-p) x y (1 - x - y) \;.
\label{MF}
\end{eqnarray}
A trivial solution of these equations is the all-centrist consensus
state, i.e., $(x,y)=(0,0)$ (or equivalently $z=1$). However, the
steady state solution (see Appendix B) of these equations
with $\alpha > \frac{1}{2}$ shows the existence of a critical point,
\begin{equation}
p_c=\frac{3(2\alpha-1)}{8+3(2\alpha-1)} \;,
\end{equation}
such that for $p < p_c$ the system exhibits two non-trivial fixed points as well:
\begin{equation}
(x,y) = \left(\frac{1}{4}+\frac{1}{4}\sqrt{1-\frac{8p}{3(2\alpha-1)(1-p)}}\right)(1,1) \;,
\end{equation}
which is a metastable fixed point and
\begin{equation}
(x,y) = \left(\frac{1}{4}-\frac{1}{4}\sqrt{1-\frac{8p}{3(2\alpha-1)(1-p)}}\right)(1,1) \;,
\end{equation}
which is a saddle point (unstable fixed point). 
In the other scenario, when $\alpha < \frac{1}{2}$, the system already has the tendency to move towards
an all-centrist consensus state, hence only the trivial fixed point $(x,y)=(0,0)$ exists.

In this paper we focus on the case of $\alpha > \frac{1}{2}$ in which the
system has the tendency to become polarized and the external
influence field is required to prevent this polarization.
In this regime, due to the competition between balancing and influencing forces, the densities
fluctuate around the metastable point and the system is trapped for
exponentially long times before unlikely large fluctuation moves it
to the absorbing state. Here all fixed points lie on the line $y=x$
as shown in Fig.~\ref{phase} and any asymmetry in $x$ and $y$ decays
exponentially fast~\cite{KornissJSP,Haken} (see Appendix B). The trajectories shown in Fig.~\ref{phase} are exact
only in the thermodynamic limit (fully-connected network in the limit of $N\to\infty$).
\begin{figure}[!htbp]
\centerline{
\epsfxsize=8 cm
\epsfclipon
\epsfbox{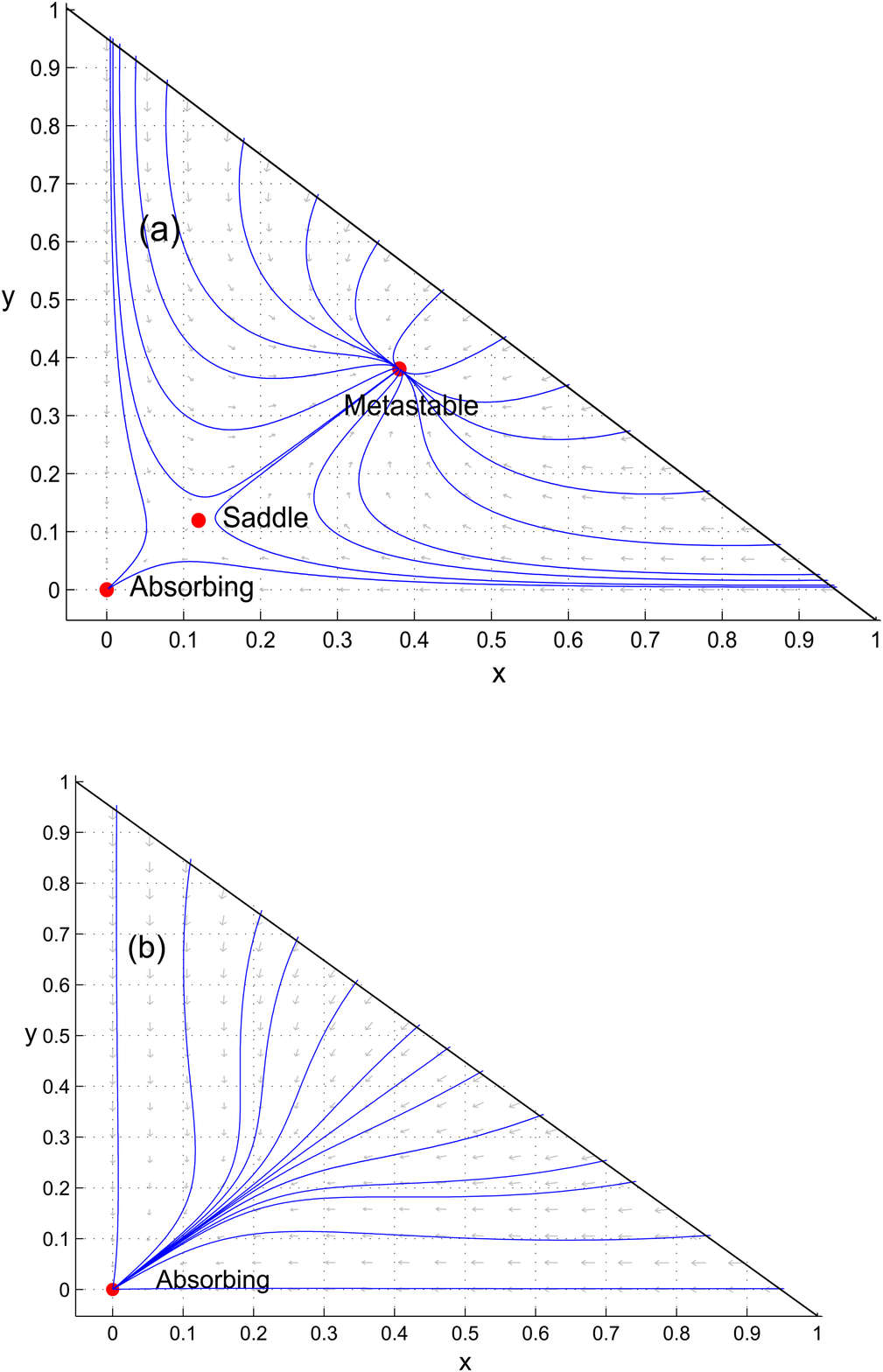} }
\caption{(Color online) Phase-space trajectories within mean-field approximation
for $\alpha$$=$$0.75$ for which $p_c$$\approx$$0.16$,
(a) for $p$$=$$0.12$$<$$p_c$, and (b) for $p=$$0.20$$>$$p_c$.}
\label{phase}
\end{figure}
It can be seen explicitly that a finite network ($N=100$) in the weak-field limit ($p<p_c$)
gets stuck in the metastable state and never crosses the saddle point whereas a fast centrist consensus
is reached when $p>p_c$ (Fig.~\ref{trajectories}).
\begin{figure}[!htbp]
\centerline{
\epsfxsize=8 cm
\epsfclipon
\epsfbox{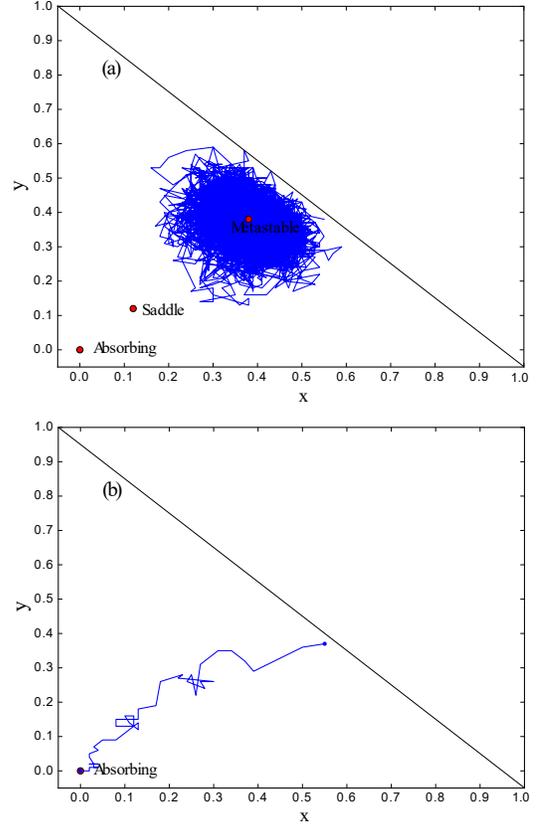} }
\caption{(Color online) Stochastic trajectories for a fully-connected network of size
$N=100$ and $\alpha=0.75$, (a) for $p$$=$$0.12$$<$$p_c$, and (b) for $p=$$0.20$$>$$p_c$.
See also Supplemental Material for animations of the above two scenarios (for the same parameters).}
\label{trajectories}
\end{figure}
It is also clear from the above that the locations of the fixed points
(roots) depend on the choice of $\alpha$ ($>\frac{1}{2}$) and $p$.
For a particular choice of $\alpha$, the two fixed points
(metastable and saddle) move closer to each other as $p$ is
increased from $0$ till they meet and annihilate each other at
$p=p_c$ (as shown in Fig.~\ref{fixed_points}). Beyond $p_c$, these
additional fixed points cease to exist and the only fixed point is
the consensus state ($x=y=0$).
\begin{figure}[!htbp]
\centerline{
\epsfxsize=9 cm
\epsfclipon
\epsfbox{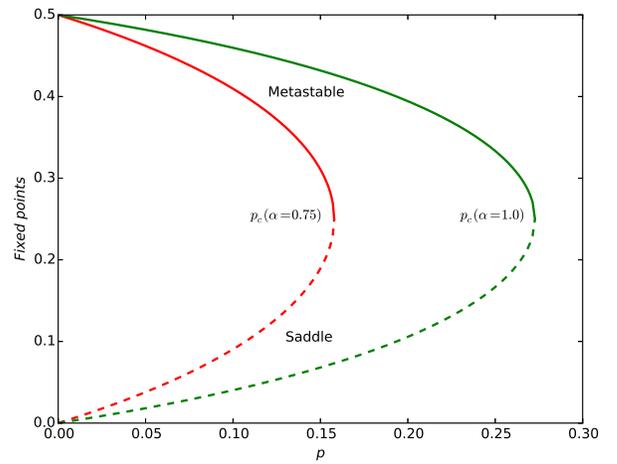}
}
\caption{(Color online) The locations ($x$, $y$ co-ordinates) of metastable (solid lines) and saddle (dashed lines) fixed points for $\alpha=0.75$ (red) and $\alpha=1.0$ (green) obtained from the solution of the rate equations. $x = y$ for all fixed points.}
\label{fixed_points}
\end{figure}
\subsection{Consensus time for finite-size networks}
An all centrist consensus state is always reached for a finite
network. Time to reach this absorbing state (consensus time $T_c$)
can be obtained by direct simulations. This approach works well for
$p > p_c$, however, for $p < p_c$ (specially when $p << p_c$ and/or
$N >> 1$), $T_c$ becomes so large that its estimation by simulation
becomes difficult if at all possible. We therefore use the
quasi-stationary (QS) approximation prescribed in~\cite{Dickman2002}
and also used in~\cite{xie2011,Dickman2004} to estimate $T_c$ in the
region $p < p_c$.

We start by introducing notation for the numbers of nodes with the given opinion, thus, $X=xN$, $Y=yN$, and $Z=zN$. Then we form the master equation that describes the time evolution of probability $P_{X,Y}$ (the probability that system has $X$ leftists and $Y$ rightists at time $t$).

\begin{eqnarray}
\frac{1}{N} \frac{dP_{X,Y}}{dt} &=&  {P}_{X-1,Y} \frac{3\alpha(1-p)(X-1)Y(N-X+1-Y)}{N(N-1)(N-2)} \nonumber \\&+& {P}_{X,Y-1} \frac{3\alpha(1-p)X(Y-1)(N-X-Y+1)}{N(N-1)(N-2)} \nonumber \\&+& {P}_{X+1,Y} \frac{3(1-\alpha)(1-p)(X+1)Y(N-X-1-Y)}{N(N-1)(N-2)} \nonumber \\&+& {P}_{X+1,Y}~p~\frac{X+1}{N} \nonumber \\&+& {P}_{X,Y+1} \frac{3(1-\alpha)(1-p)X(Y+1)(N-X-Y-1)}{N(N-1)(N-2)} \nonumber \\&+& {P}_{X,Y+1}~p~\frac{Y+1}{N} \nonumber \\&-& P_{X,Y} \frac{6(1-p)XY(N-X-Y)}{N(N-1)(N-2)} \nonumber \\&-& P_{X,Y}~p~\frac{X+Y}{N} \;.
\label{master}
\end{eqnarray}

Within the triangular region (bounded by $0 \le X \le (N-Y)$ and $0 \le Y \le (N-X)$), the transitions allowed from a state ($X$, $Y$) are to states ($X \pm 1$, $Y$) or ($X$, $Y \pm 1$) with the constraint that the system stays within the bounded region. The positive and negative terms on the right side of the master equation contribute to the net flow of probability into and out of the state ($X$,$Y$), respectively. A factor of $\frac{1}{N}$ on the left-hand side appears because a microscopic step of transition from initial state to the final state takes place in a time interval $1/N$. $\frac{6~X~Y~(N-X-Y)}{N(N-1)(N-2)}$ is the density of unbalanced triangles at any given time for a fully-connected network.

The QS distribution of occupation probabilities is given by
$\tilde{P}_{X,Y}=P_{X,Y}(t)/P_S(t)$ where $P_S(t)$ is the survival
probability. Under the QS hypothesis, the survival probability
decays exponentially, governed by
\begin{equation}
\frac{dP_S(t)}{dt} =  - P_S(t) \tilde{Q}_0 \;,
\label{qs}
\end{equation}
where $\tilde{Q}_0 = p [\tilde{P}_{1,0} + \tilde{P}_{0,1}]$ measures
the flow of probability into the absorbing state $(0,0)$. The underlying idea of QS hypothesis
is that the occupation probability distribution conditioned on survival (over all $(X, Y)$ except the absorbing state)
is stationary. Therefore,
\begin{equation}
\frac{dP_{X,Y}}{dt} = \tilde{P}_{X,Y} \frac{dP_S(t)}{dt} \;.
\label{dPdt}
\end{equation}

We plug in $P_{X,Y}$ in terms of $\tilde{P}_{X,Y}$ into the master equation to
obtain the QS distribution,
\begin{equation}
\tilde{P}_{X,Y} =  \frac{\tilde{Q}_{X,Y}}{W_{X,Y}-\tilde{Q}_0} \;,
\label{qsp}
\end{equation}
where $W_{X,Y}=\frac{6(1-p)XY(N-X-Y)}{(N-1)(N-2)}+p(X+Y)$, and
\begin{eqnarray}
\tilde{Q}_{X,Y} &=&  \tilde{P}_{X-1,Y} \frac{3\alpha(1-p)(X-1)Y(N-X+1-Y)}{(N-1)(N-2)} \nonumber \\&+& \tilde{P}_{X,Y-1} \frac{3\alpha(1-p)X(Y-1)(N-X-Y+1)}{(N-1)(N-2)} \nonumber \\&+& \tilde{P}_{X+1,Y} \frac{3(1-\alpha)(1-p)(X+1)Y(N-X-1-Y)}{(N-1)(N-2)} \nonumber \\&+& \tilde{P}_{X+1,Y}~p~(X+1) \nonumber \\&+& \tilde{P}_{X,Y+1} \frac{3(1-\alpha)(1-p)X(Y+1)(N-X-Y-1)}{(N-1)(N-2)} \nonumber \\&+& \tilde{P}_{X,Y+1}~p~(Y+1) \nonumber \;.
\label{tildeq}
\end{eqnarray}

\begin{figure}[!htbp]
\centerline{
\epsfxsize=9 cm
\epsfclipon
\epsfbox{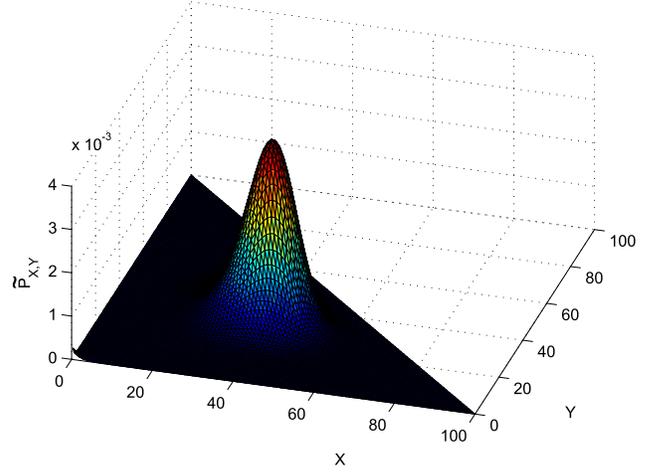} }
\caption{(Color online) The QS distribution with $a$$=$$0.5$ for $N$$=$$100$, $\alpha$$=$$0.75$, $p$$=$$0.12$.
For these parameters, $p_c$$\approx$$0.16$.}
\label{qsdist}
\end{figure}

Starting from an arbitrary distribution $\tilde{P}_{X,Y}^0$, an
asymptotic QS distribution $\tilde{P}_{X,Y}$ can be obtained by the
iteration: $\tilde{P}_{X,Y}^{i+1} = a~\tilde{P}_{X,Y}^i +
(1-a)~\frac{\tilde{Q}_{X,Y}^{i}}{ W_{X,Y}^{i} - \tilde{Q}_0^{i}}$,
where $0 \leq a \leq 1$ is an arbitrary parameter~\cite{xie2011}.
The QS distribution for a particular system size $N=100$
(fully-connected), $p=0.12$, and $\alpha=0.75$ with parameter
$a=0.5$ is shown in Fig.~\ref{qsdist}. In
this case a satisfactory convergence was obtained in $40000$
iterations. As expected from the mean-field analysis (for $\alpha=0.75$, the metastable fixed point
$(x, y) = (0.38, 0.38)$) the distribution peaks around $(X, Y)=(38,
38)$.

Once the desired QS distribution is
obtained, the mean consensus time $T_c$ is computed from the decay
rate of the survival probability,

\begin{equation}
T_c \simeq \frac{1}{p~[\tilde{P}_{1,0}+\tilde{P}_{0,1}]} \;.
\label{tc}
\end{equation}

\begin{figure}[!htbp]
\centerline{
\epsfxsize=8 cm
\epsfclipon
\epsfbox{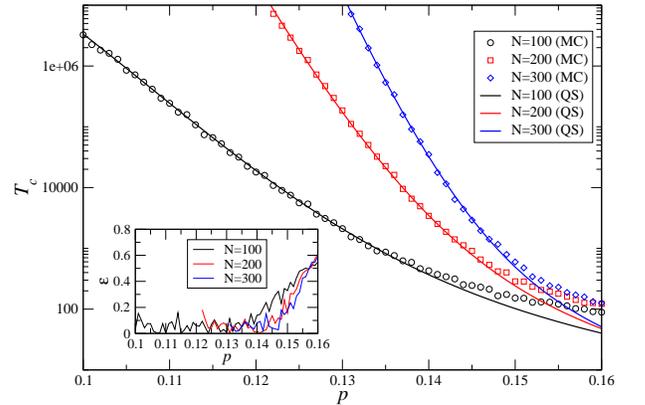}
}
\caption{(Color online) Consensus time $T_c$ obtained from the QS distribution and by direct MC simulations for $\alpha$$=$$0.75$ ($p_c$$\approx$$0.16$).}
\label{qs_mc_75}
\end{figure}

\begin{figure}[!htbp]
\centerline{
\epsfxsize=8 cm
\epsfclipon
\epsfbox{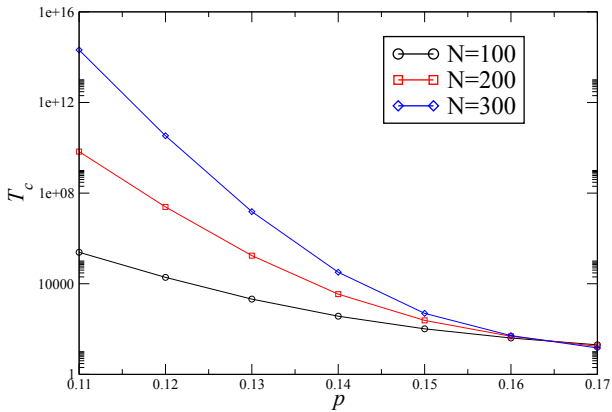}
}
\caption{(Color online) Consensus time $T_c$ computed by the QS approximation as a function of $p$ in the weak-field regime for $\alpha$$=$$0.75$ ($p_c$$\approx$$0.16$).}
\label{tc_p_qs}
\end{figure}

\begin{figure}[!htbp]
\centerline{
\epsfxsize=8 cm
\epsfclipon
\epsfbox{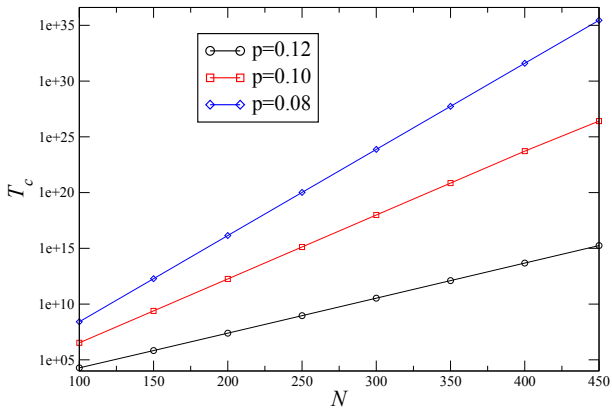}
}
\caption{(Color online) Consensus time $T_c$ computed by the QS approximation as a function of system size $N$ in the weak-field regime for $\alpha$$=$$0.75$ ($p_c$$\approx$$0.16$).}
\label{tc_n_qs}
\end{figure}

We compare $T_c$ obtained from the QS approximation to that
obtained by direct Monte Carlo (MC) simulations in the region of $p< p_c$ where $T_c$
could be easily obtained by both methods for the respective system sizes [Fig.~\ref{qs_mc_75}].
One can see that there is a good agreement between the two methods across many system sizes and the
agreement is getting better as $p$ is decreased below $p_c$ as shown in
Fig.~\ref{qs_mc_75}. The change in the relative error with respect to $p$ and $N$,
 $\epsilon=\frac{\lvert T_c(QS)-T_c(MC)\rvert}{T_c(MC)}$ can be seen in the inset.
Figure~\ref{tc_p_qs} shows $T_c$ obtained by the QS approximation as a function
of $p$ for the entire range of $p$ considered for all system sizes,
where MC simulations become prohibitive to estimate $T_c$. The
consensus time $T_c$ shows an exponential scaling with $N$ for $p <
p_c$ as shown in Fig.~\ref{tc_n_qs}.

\begin{figure}[!htbp]
\centerline{
\epsfxsize=8 cm
\epsfclipon
\epsfbox{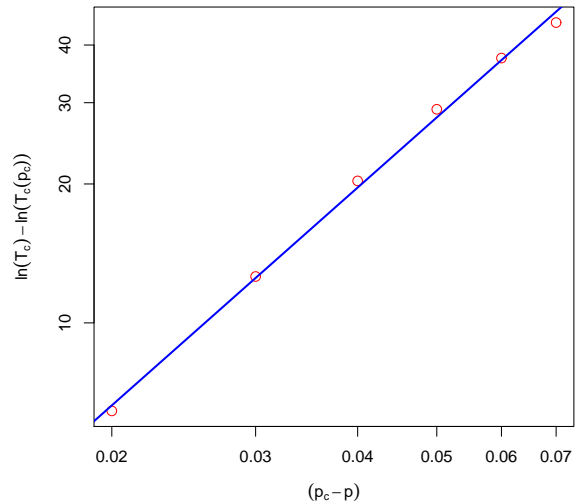}
}
\caption{(Color online) $\ln(T_c) - \ln(T_c(p_c))$ as a function of $(p_c - p)$ on a log-log scale. The exponent $\nu$ is given by the slope of the fitted line ($N=300$, $\alpha=0.75$, $p_c$$\approx$$0.16$, $\nu \approx 1.57$).}
\label{beta}
\end{figure}

In order to obtain the dependence of $T_c$ on $p$, we assume a relation common in systems with tipping points and barrier crossing \cite{xie2011,xie2012},
\begin{equation}
T_c = f(N) \exp[\beta(p) N] \;,
\end{equation}
where  $f(N)$ is increasing slower than exponential with $N$ and
\begin{equation}
\beta(p) \sim \lvert p_c - p\rvert^\nu \;.
\end{equation}
With $T_{c}(p_c)=f(N)$, we have
\begin{equation}
\ln(T_c) - \ln(T_c(p_c)) = \beta(p) N \;.
\end{equation}
As can be seen from Fig.~\ref{beta}, the growth rate approximately follows the scaling behavior $\beta \sim \lvert p_c - p\rvert^\nu$ with the measured exponent $\nu \approx 1.57$.
\section{Low-dimensional networks}
\begin{figure}[!htbp]
\centerline{
\epsfxsize=8 cm
\epsfclipon
\epsfbox{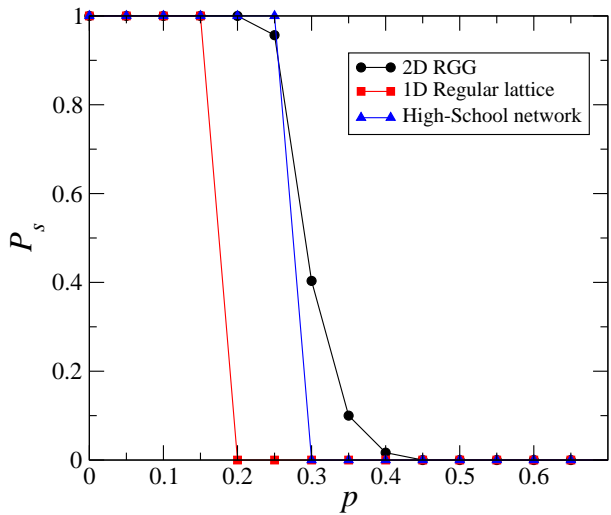}
}
\caption{(Color online) Survival probability at time $t$$=$$5000$, as a function of $p$ for $\alpha$$=$$1.0$ 
for a 1D regular lattice (with degree $k$$=$$10$), a 2D RGG (with $\langle k\rangle$$=$$10$)
with $N$$=$$1000$, and for a high-school friendship network with $\langle k\rangle$$\approx$$6$ and $N=921$.}
\label{ps_p}
\end{figure}

For sparse low-dimensional networks, the mean-field analysis does not hold and
QS approach is difficult to formulate. Hence we rely solely on MC
simulations. Specifically, we look at the survival probability $P_s$ (for a fixed cutoff time $t=5000$) for a 2D random geometric graph
(RGG) \cite{Dall_PRE2002} with $\langle k\rangle=10$ and a 1D regular lattice with each
node having a degree $k = 10$. We start with a polarized initial
state ($x=0.5, y=0.5$). The simulation results indicate the
existence of a critical point $p_c$ at which the survival
probability undergoes an abrupt transition~\cite{singh2012} as shown
in Fig.~\ref{ps_p}. We choose these particular spatial embeddings
because presence of local clustering ensures a significant number of
triangles in the network and the model requires the presence of
triangles for the balance dynamics to take place. For networks with
relatively low clustering coefficient (e.g. ER, BA networks), the
dynamics would be heavily dominated by external influence.
To examine the dynamics on a real-world network, we also simulated the dynamics, starting from the same initial conditions (randomly assigning
the opinions in the initial state of the system),
on the giant component of a high-school friendship network from the {\it Add Health} data set~\cite{hs}.
The critical point in $p_c$ is shown to exist in this network structure as well (Fig.~\ref{ps_p}).

\begin{figure}[!htbp]
\centerline{
\epsfxsize=7 cm
\epsfclipon
\epsfbox{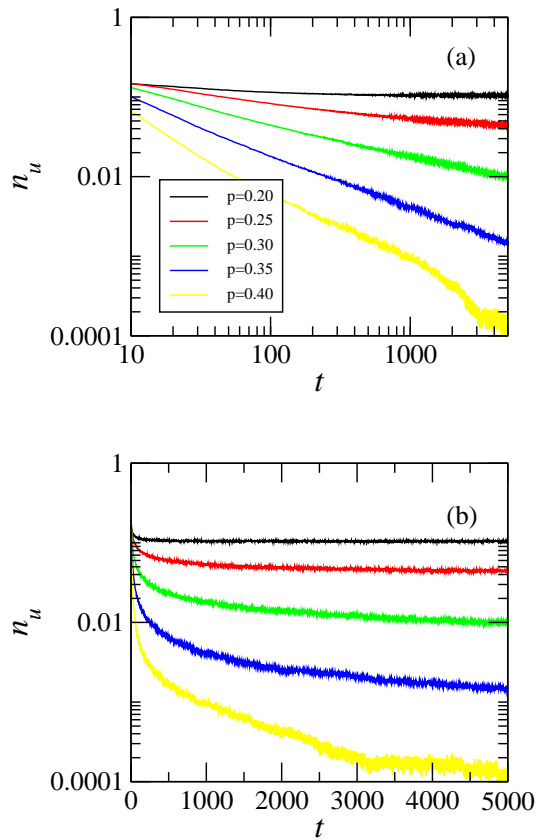}
}
\caption{(Color online) Average fraction of unbalanced triangles in the 2D RGG with $N=1000$, $\langle k\rangle=10$, and $\alpha=1.0$ (a) on a log-log scale and (b) on a semi-log scale. The initial population densities are $x=0.5, y=0.5$.}
\label{nu_rgg}
\end{figure}

\begin{figure}[!htbp]
\centerline{
\epsfxsize=7 cm
\epsfclipon
\epsfbox{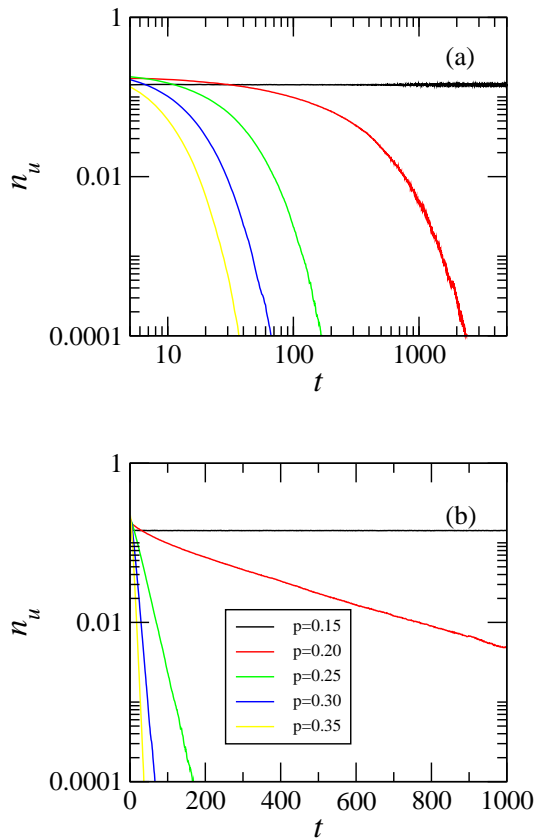}
}
\caption{(Color online) Average fraction of unbalanced triangles in the 1D regular lattice with $N=1000$, $k=10$, and $\alpha=1.0$ (a) on a log-log scale and (b) on a semi-log scale. The initial population densities are $x=0.5, y=0.5$.}
\label{nu_1d}
\end{figure}

\begin{figure}[!htbp]
\centerline{
\epsfxsize=6 cm
\epsfclipon
\epsfbox{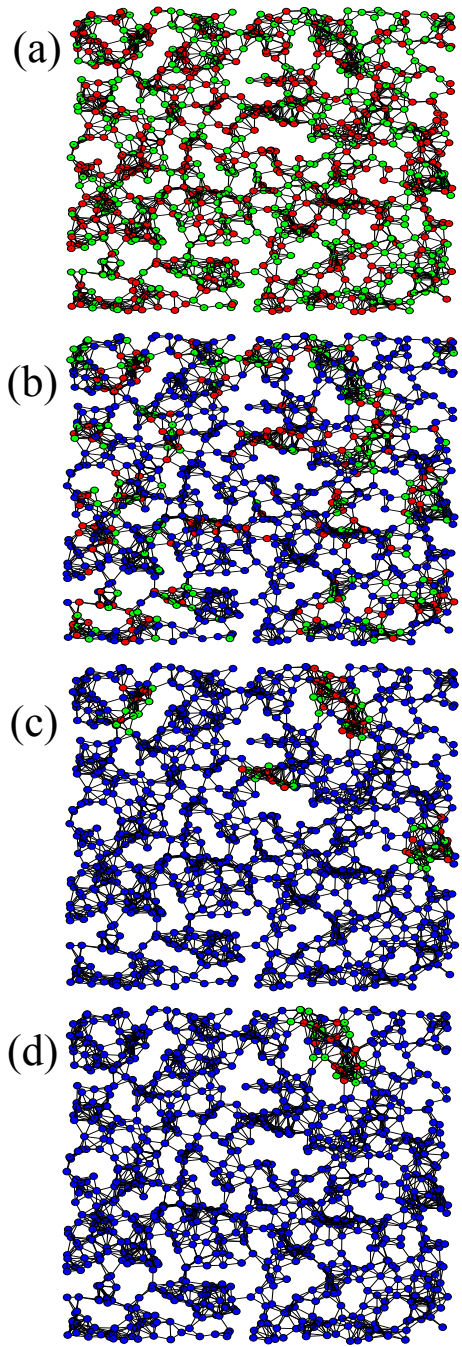}
}
\caption{(Color online) Time of evolution (single-run) of the system. Green, red, and blue nodes correspond to leftists, rightists, and centrists respectively. The network structure is 2D RGG with $\langle k\rangle=10$, $N=1000$, $\alpha=1.0$, and $p=0.3$. System is initialized with $x=0.5, y=0.5$. (a) $t=0$, (b) $t=10$, (c) $t=100$, and (d) $t=500$, where $t$ is the number of time steps.}
\label{rgg_snapshots}
\end{figure}

The simulation results (Fig.~\ref{ps_p}) indicate that the critical
point for 2D RGG is significantly higher than that for 1D lattice.
For the same choice of $\alpha=1$, the
critical point in the case of fully-connected network is
$p_c\approx0.27$ (between the values of the 1D and 2D systems).
Simulation results plotted in Figs.~\ref{nu_rgg}(a) and (b) show that the decay of fraction of
unbalanced triangles in the network ($n_u$) is governed by the power-law in the case of 2D RGG but it decays exponentially (ignoring the
transience) in 1D regular lattice as shown in Figs.~\ref{nu_1d}(a)
and (b). In the case of 2D RGG, there are frustrated domains
that are long-lived (shown in Fig.~\ref{rgg_snapshots}), whose existence can be causal or symptomatic  
to the slow convergence. Particular network
structure and spatial correlations (among nodes) become
significantly important in defining the dynamics of low dimensional
systems and a quantitative analysis becomes mathematically
challenging.

\section{Summary}
In summary, we presented a framework that models social imbalance
arising from individual opinions in the simplest manner possible,
and observe its counteracting effect on an externally influencing
field. We found that there exists a critical value $p_c$, above
which the only fixed point is the centrist consensus state, and
below which a metastable fixed point of the system emerges. We
demonstrated how the competition between balance and influence can
lead the system to metastability. Using a semi-analytical approach
(QS approximation), we estimated the consensus times which show good
agreement with simulation results. Additionally, employing
simulations, we demonstrated that this critical behavior is also
seen in sparse networks.

\section{Acknowledgements}
This work was supported in part
by the Army Research Laboratory under Cooperative Agreement Number W911NF-09-2-0053 (the ARL Network Science CTA),
by the Office of Naval Research Grant Nos.~N00014-09-1-0607 and  N00014-15-1-2640,
by the Army Research Office grant W911NF-12-1-0546,
by the European Commission under the 7th Framework Programme, Grant Agreement Number 316097 [ENGINE],
and by the National Science Centre, Poland, the decision no. DEC-2013/09/B/ST6/02317.
The views and conclusions contained in this document are those of
the authors and  should not be interpreted as representing the
official policies either expressed or implied of the Army Research
Laboratory or the U.S. Government.

\appendix
\section{The $p = 0$ case}
\begin{figure}[!htbp]
\centerline{
\epsfxsize=7.5 cm
\epsfclipon
\epsfbox{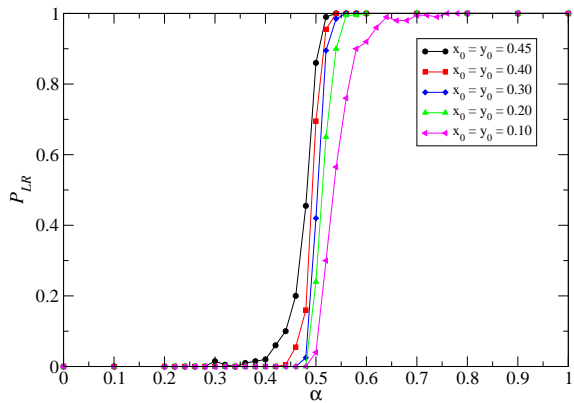} 
}
\caption{(Color online) The probability $P_{LR}$ as a function of $\alpha$ for $N=100$ (fully-connected) and different starting points ($p=0$).}
\label{plr}
\end{figure}

\begin{figure}[!htbp]
\centerline{
\epsfxsize=7 cm
\epsfclipon
\epsfbox{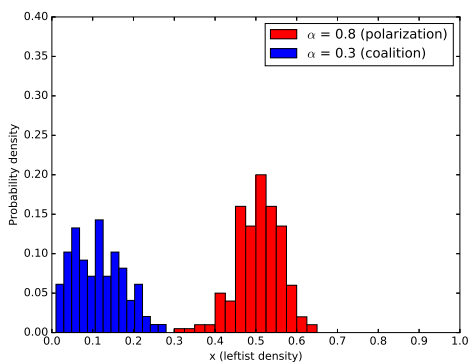} 
}
\caption{(Color online) The histogram of leftist density $x$ when the system ends up in a leftist-rightist polarization or centrist-leftist coalition state for $\alpha=0.8$ and $\alpha=0.3$. The network is fully-connected ($N=100$ and $p=0$).}

\label{hist}
\end{figure}

In this appendix we discuss the $p=0$ case of the model (case with no external influence). When $p=0$, the 
rate equations for the densities $x$ and $y$ can be written as
\begin{eqnarray}
\frac{dx}{dt} &=& 3 (2\alpha-1) x y (1 - x - y)  \nonumber  \\
\frac{dy}{dt} &=& 3 (2\alpha-1) x y (1 - x - y) \;.
\end{eqnarray}
In this case, the entire boundary of the triangular phase space becomes absorbing because structural balance is achieved as soon as either of the $x$, $y$, or $z$ variables becomes zero and from that point the system does not evolve.
The type of steady-state (in other other words, which of the three boundaries is hit by the system) of this system depends on the choice of $\alpha$. For $\alpha < \frac{1}{2}$, the system moves towards an all-centrist consensus till it eventually hits either $x = 0$ (centrist-leftist coalition) or $y = 0$ (centrist-rightist coalition) boundary. However, for $\alpha > \frac{1}{2}$ (in the absence of external field, $p = 0$) the system tends to more radical configuration and stops evolving when $x + y = 1$ (or reaches the long side of the triangular phase space). After gaining some insight in this case, we performed stochastic simulations. The probability for the system to end up in a polarized state $P_{LR}$ is shown in Fig.~\ref{plr} (with the complementary probability $P_{CE} = 1 - P_{LR}$, the system settles in a coalition state). 
Starting from an equal density initial state ($x_0 \approx y_0 \approx z_0 \approx 0.33$), we also look at the composition of the final state when the system reaches a polarized ($\alpha = 0.8$) or a coalition state ($\alpha = 0.3$). We only show the distribution of $x$ given that the system reaches a final state on the $y = 0$ or $y = 1 - x$ boundary (Fig.~\ref{hist}). $y$ has the same distribution along $x = 0$ and $y = 1 - x$ boundaries due to symmetry. Starting from an arbitrary state ($x_0 >0, y_0 > 0, z_0 > 0$), the system never reaches a pure consensus because a structurally balanced configuration is always reached before reaching a pure consensus state and the system freezes in that state. The consensus in this case is observed only if the initial state itself is a consensus state and the system remains in that state.  
\section{Steady-state solution of the rate equations}
The rate equations for the leftist and rightist densities under the mean-field assumption are
\begin{eqnarray}
\frac{dx}{dt} &=& -p x  +  3 (2\alpha-1) (1-p) x y (1 - x - y)  \nonumber  \\
\frac{dy}{dt} &=& -p y  +  3 (2\alpha-1) (1-p) x y (1 - x - y) \;.
\label{mf_app}
\end{eqnarray}
By adding and subtracting the above equations and introducing a new
set of variables $u = (x + y)$ and $v = (x - y)$, one can
immediately see that
\begin{equation}
\frac{dv}{dt} = -p v \;,
\end{equation}
yielding $v \sim \exp(-pt)$, which means that $v \to 0$
exponentially fast. Therefore, we can assume that $x \approx y$
which allows us to analyze the system in terms of single-variable
equation for the ``slow" mode $u$ (and $x = y = u/2$),
\begin{equation}
\frac{du}{dt} = -p u  +  6 (2\alpha-1) (1-p) \frac{u^{2}(1-u)}{4} \;,
\end{equation}
which can be solved for stead-state $\frac{du}{dt} = 0$. A trivial
solution of this equation is $u = 0$, which is the absorbing state
($x = 0$, $y = 0$). Additional roots are the solutions of the
quadratic equation
\begin{equation}
u^{2} - u + \frac{2p}{3(1-p)(2\alpha-1)} = 0 \;,
\end{equation}
and are given by
\begin{equation}
u = \frac{1}{2} \pm \frac{1}{2} \sqrt{1-\frac{8p}{3(1-p)(2\alpha-1)}} \;.
\end{equation}
These solutions make sense only when $\alpha > 1/2$, otherwise the
solution will lie outside the feasible domain [$(x + y) \le 1$]. For
$\alpha > 1/2$, we obtain a critical point,
\begin{equation}
p_c = \frac{3(2\alpha-1)}{8+3(2\alpha-1)} \;,
\end{equation}
such that the roots are real and positive for $p < p_c$. Thus, in
terms of $x$ and $y$ the two roots (other than the absorbing state)
are
\begin{eqnarray}
x = y &=& \frac{1}{4} + \frac{1}{4} \sqrt{1-\frac{8p}{3(1-p)(2\alpha-1)}}~~~~~{\rm (metastable)} \nonumber  \\
x = y &=& \frac{1}{4} - \frac{1}{4} \sqrt{1-\frac{8p}{3(1-p)(2\alpha-1)}}~~~~~{\rm (saddle)}     \;.
\end{eqnarray}



\end{document}